\documentclass[aps,prmaterials,preprint,floatfix,superscriptaddress,]{revtex4-1}

\usepackage[utf8]{inputenc}
\usepackage{array}
\usepackage{graphicx}
\usepackage{amsmath}
\usepackage{amsfonts}
\usepackage{color}
\usepackage{dcolumn}
\usepackage{bm}

\begin{document}

\title{Tunable Topology and Berry Curvature Dipole in Transition Metal Dichalcogenide Janus Monolayers}
\author{Nesta Benno Joseph}
\affiliation{Solid State and Structural Chemistry Unit, Indian Institute of Science, Bangalore 560012, India.}
\author{Saswata Roy}
\affiliation{Undergraduate Programme, Indian Institute of Science, Bangalore 560012, India.}
\author{Awadhesh Narayan}
\email{awadhesh@iisc.ac.in}
\affiliation{Solid State and Structural Chemistry Unit, Indian Institute of Science, Bangalore 560012, India.}

\date{\today}

\begin{abstract}
Janus transition metal dichalcogenides, with intrinsic mirror asymmetry, exhibit a wide array of interesting properties. In this work, we study Janus monolayers derived from WTe$_2$ using first-principles and tight-binding calculations. We discover that WSeTe and WSTe are topologically trivial, in contrast to the parent quantum spin Hall insulator WTe$_2$. Motivated by the growing interest in non-linear Hall effect, which also requires asymmetric structures, we investigate the Berry curvature and its dipole in these Janus systems and find that they exhibit strikingly large values of Berry curvature dipole, despite being in the topologically trivial phase. We track down the origin of this behaviour and put forth a low-energy massive Dirac model to understand the central features of our \textit{ab inito} computations. Our predictions introduce Janus monolayers as promising new platforms for exploring as well as engineering non-linear Hall effect.
\end{abstract}

\maketitle


\section{\label{sec:intro}Introduction}
 
Two dimensional materials have been a prime focus of research ever since the discovery of graphene~\cite{novoselov2004electric} and identification of its novel properties such as high carrier mobility~\cite{BOLOTIN2008351}, optical transparency~\cite{graphene_optical} and high electrical and thermal conductivity~\cite{balandin2008superior}.
Transition metal dichalcogenides (TMDCs), with an X-M-X sandwich structure, M being a transition metal within group III - VI and X being either S, Se or Te, are one of the most intensively studied two-dimensional materials~\cite{manzeli20172d}. 
They exhibit a staggeringly wide array of electronic and structural properties, and are potential candidates for diverse applications~\cite{chhowalla2013chemistry,bhimanapati2015recent,manzeli20172d}.
Depending on the M-X coordination, TMDCs exist in three different phases - trigonal prismatic coordinated 2H, octahedral coordinated 1T and the distorted octahedral coordinated 1T'.
Among the three phases, the ground state configuration of each TMDC depends on its composition, with the stable 2H MoS$_2$ most studied as the representative member of the group~\cite{ganatra2014few}.
This is because the semiconducting material was found to display an indirect to a direct bandgap transition when thinned down from bulk to monolayer, as a result of quantum confinement~\cite{splendiani2010emerging}.
Most of the members display ground state stability in either 2H or 1T coordination, the other phases being a higher energy or a meta-stable state, with the exception of WTe$_2$.
WTe$_2$ is the only known TMDC to display ground state stability in its 1T' phase and exists as a Weyl semimetal~\cite{soluyanov2015type} and quantum spin Hall insulator~\cite{Qian@tmdcQSH_science} in its bulk and monolayer forms, respectively.



The exploitation of electron spin for the development of spintronic devices requires materials having an enhanced spin-orbit coupling interaction along with broken inversion symmetry, leading to a large Rashba effect~\cite{manchon2015new}.
Monolayer 2H phase of TMDCs became a viable candidate, but the lack of Rashba splitting due to the out-of-plane mirror symmetry was a hurdle to overcome.
Janus monolayers of TMDCs, first proposed by Cheng \textit{et al.}~\cite{cheng2013spin}, became a possible solution. 
In Janus TMDCs, one layer of chalcogen in the MX$_2$ sandwich monolayer is replaced by another, leading to an asymmetric MXY structure.
It was predicted that the broken out-of-plane symmetry in these systems could lead to distinct and novel properties in the Janus monolayers, compared to their parent materials~\cite{shi2018mechanical,wang2018intriguing}.
Different experimental strategies have been devised to synthesise such asymmetric structures, with the first Janus monolayer TMDC MoSSe synthesised by Zhang \textit{et al.}~\cite{zhang2017janus} and Lu \textit{et al.}~\cite{lu2017janus}.
After the synthesis of MoSSe, other two-dimensional Janus structures were also predicted to exist. 
The half-metallic, ferromagnetic FeXY~\cite{li2021two}, TMDC oxides WXO~\cite{varjovi2021janus}, In$_2$SSe, the Janus analogue of binary compounds InSe and InS ~\cite{kandemir2018janus} and Janus structures derived from metal monochalcogenides~\cite{liu2021tuning} are a few examples.
The different Janus materials show several intriguing properties~\cite{li2018recent,yagmurcukardes2020quantum,ju2021two}, including the aforementioned large Rashba effect~\cite{hu2018intrinsic}, tunable Dzyaloshinskii-Moriya interactions~\cite{liang2020very}, piezoelectricity and novel valleytronic effects~\cite{zhang2019first,idrees2019optoelectronic,nandi2021group,ahammed2020ultrahigh}.
Another major application of these materials is in water splitting and hydrogen evolution reactions, as these are predicted to exhibit enhanced photocatalytic activity~\cite{er2018prediction,ju2020janus,peng2019two}.\\


In a parallel, yet distinct, thread of developments, non-linear generalizations of the Hall effect have been proposed. Following the prediction of Sodemann and Fu, the first moment of the Berry curvature, termed Berry curvature dipole (BCD), has been identified as an important quantity responsible for non-linear Hall effects~\cite{PhysRevLett.115.216806}. 
Such a non-linear Hall effect can also be generated in systems preserving time reversal (TR) symmetry, in contrast to the linear Hall effect ~\cite{PhysRevLett.115.216806,du2021perspective,ortix2021nonlinear}. 
Diverse set of materials have been predicted to show finite BCD and the resulting second order Hall effect, including two-dimensional materials ~\cite{PhysRevB.92.235447,PhysRevB.98.121109,zhang2018electrically,joseph2021topological,PhysRevB.103.L201202}, and Dirac and Weyl semimetals ~\cite{PhysRevLett.121.266601,PhysRevB.102.024109,PhysRevLett.125.046402,PhysRevB.97.041101,roy2021non}, strained bilayer and monolayer graphene \cite{PhysRevLett.123.196403} and Rashba systems~\cite{PhysRevLett.121.246403}. 
Significant advances have also been made in the experimental detection of such Hall signals in several material platforms, starting with few-layer WTe$_2$~\cite{Ma2019,shvetsov2019nonlinear,dzsaber2021giant,huang2020giant,kiswandhi2021observation,he2021quantum,kumar2021room}. 
Building on these rapid developments, higher order Hall responses have also been predicted~\cite{2106.04931,2012.15628}, and very recently realized experimentally~\cite{Lai2021}.\\

Motivated by these advances, in this work, we investigate Janus monolayers derived from 1T' WTe$_2$ using first-principles density functional calculations. 
As mentioned above, unlike other TMDCs, the stable phase of WTe$_2$ is 1T'. 
Previous studies on Janus WXY concentrated on the 2H phase of the material~\cite{kandemir2018bilayers,ju2020janus}, for example, with application as photo catalysts in water splitting reactions~\cite{ju2020janus}. 
The topologically non-trivial ground state of WTe$_2$ make this material an intriguing platform to look for tunable topology, in context of this study, as a result of replacing one layer of Te with its lighter congeners.
We show that WSeTe and WSTe are topologically trivial, in contrast to WTe$_2$, which is a quantum spin Hall insulator. 
Using Wannier function based tight binding models, we find that although the edge states in the three systems appear very similar, a closer examination reveals their topologically distinct nature. 
We investigate the BCD in these systems and find that all of them exhibit large values of BCD, which are tunable in the Janus structures. 
Remarkably, the BCD shows a pronounced peak near the Fermi level in the WSTe system, owing to a reduced band gap which results in a larger Berry curvature. 
We supplement our findings with an effective low-energy massive tilted Dirac model, which captures the essential features of our first-principles results. Our studies place Janus monolayers as promising platforms to engineer non-linear Hall effects.
We hope that our work motivates further investigations of topology and Berry curvature dipole in transition metal dichalcogenide Janus structures.\\


\section{\label{sec:methods} Computational Details}

Our first-principles calculations were carried out based on the density functional theory (DFT) framework as implemented in the {\sc quantum espresso} code~\cite{QE-2017,QE-2009}. 
A kinetic energy cut-off of $60$, $80$ and $55$ Ry were considered for WTe$_2$, WSeTe and WSTe respectively, using the ultrasoft pseudopotentials~\cite{PhysRevB.41.7892} to describe the core electrons, including spin-orbit coupling interactions. We used the Perdew-Burke-Ernzerhof (PBE) form for the exchange correlation functional~\cite{perdew1996generalized}.
The Janus structures were derived starting from WTe$_2$, by replacing the top layer Te with Se/S atoms and fully relaxing the resulting structures until forces on each atom were less than at least $10^{-4}$ eV/\AA.
The Brillouin zone was sampled over a uniform $\Gamma$-centered $k$-mesh of $12\times8\times1$.
The monolayers were modelled with at least 15 \AA~vacuum to avoid any spurious interaction between the periodic images. \\

To study the topological properties, maximally localized Wannier functions (MLWFs) were computed to derive a tight-binding model from the \textit{ab initio} calculations, with W $d$ orbitals, Te/Se $p$ orbitals, and S $s$ and $p$ orbitals as basis, using the {\sc wannier90} code~\cite{mostofi2014updated}. 
Two methods were employed to characterise the topological nature of the monolayers -- direct calculation of the Z$_2$ topological invariant and a computation of edge states. 
The calculation and further analysis of edge spectrum was performed using the WannierTools code~\cite{WU2017}, employing the method of calculating surface Green's function using the iterative Green's function approach~\cite{PhysRevB.28.4397,Sancho_1984,Sancho_1985}. 
A computationally feasible method to calculate the $Z_2$ topological invariant was proposed by Rui \textit{et al.}~\cite{Z2_Rui} and Soluyanov \textit{et al.}~\cite{soluyanov2011computing}, using the notion of Wannier charge centers (WCCs).
The Wannier functions (WFs) associated with any unit cell are given by

\begin{equation}
    \big|R_n\big \rangle = \frac{1}{2\pi} \int_{-\pi}^{\pi} dk~e^{-ik(R-x)}~\big|u_{nk}\big \rangle,
\end{equation}

where $\big|u_{nk}\big \rangle$ is the periodic part of the Bloch state of band $n$ at momentum $k$. 
The expectation value of the position operator in the state $\big| 0_n \big\rangle$ corresponding to WFs in the unit cell $R = 0$ gives the WCCs. 
The evolution of these WCCs of the MLWFs, along the TR invariant plane $k_z = 0$ is used to characterise the material as topologically trivial or non-trivial. \\

For non-centrosymmetric systems, Hall-like currents have been predicted to be present, even when the system preserves TR symmetry. 
This Hall effect, albeit in the non-linear regime, arise due to the presence of BCD.
For two-dimensional systems, the Berry curvature, given by

\begin{equation}
    \Omega_z^n (\textbf{k}) =  2i\hbar^2 \sum_{m \neq n} \frac{\langle n|\Hat{v}_x|m\rangle \langle n|\Hat{v}_y|m\rangle}{(\epsilon_n - \epsilon_m)^2},
\end{equation}

has only an out-of-plane ($z$) component present. 
Here, $\epsilon_m$ and $|m\rangle$ are the $m$-th energy eigenvalue and eigenvector of the Hamiltonian, $\hat{v}_x$ and $\hat{v}_y$ are the velocity operators along $x$ and $y$ respectively, and $\hbar$ is the reduced Planck's constant.
BCD is a measure of the first order moment of the Berry curvature.
For two-dimensional materials, BCD behaves as a pseudo-vector and is given by~\cite{PhysRevLett.115.216806}

\begin{equation}
    D_{az} = \int_{k}~f_n^0(\textbf{k})~\frac{\partial\Omega^n_z}{\partial k_a}d\textbf{k},
\end{equation}

with $f_n^0$(\textbf{k}) being the equilibrium Fermi-Dirac distribution and $a\in\{x,y\}$ denotes the direction. 
The non-zero components of BCD are further determined by the symmetries of the system.
We used the Wannier-based tight-binding Hamiltonian to compute the Berry curvature and BCD associated with it using the {\sc wannier-berri} code~\cite{tsirkin2021high}.


\section{\label{sec:results} Results and Discussion}

\begin{figure*}
\begin{center}
  \includegraphics[scale=0.54]{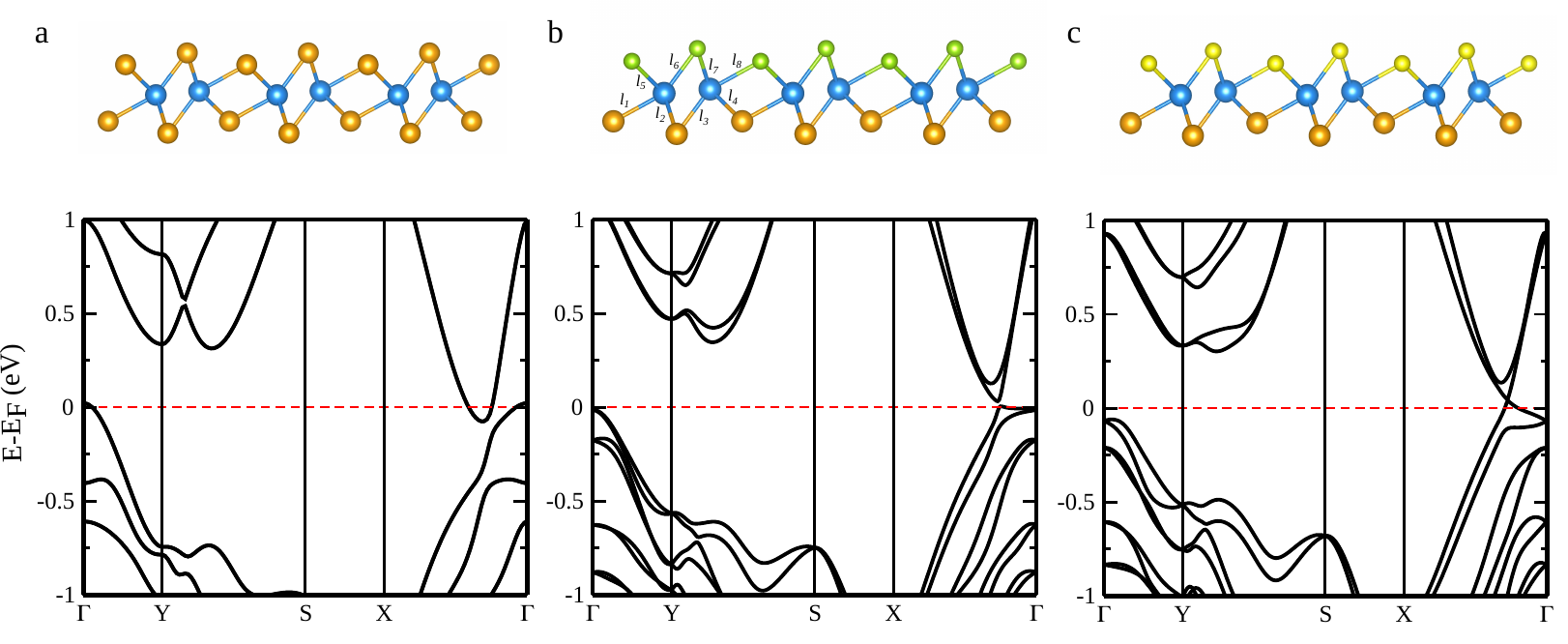}
   \caption{\textbf{Janus monolayers and their band structures.} Structure of (a) monolayer 1T' WTe$_2$ and its associated Janus structures, (b) WSeTe and (c) WSTe, along with the respective electronic bands structures. The different types of W-X bonds present are marked \textit{$l_1$ -- $l_8$}. Here, the blue spheres depict the W atoms, while the orange, green and yellow spheres represent the Te, Se and S atoms respectively. The bands split due to the strongly broken out-of-plane symmetry and the bandgap along X-$\Gamma$ reduces as one layer of Te is replaced with its lighter congeners.} \label{fig:structure}
\end{center}
\end{figure*}

\begin{table*}[]
    \centering
    \begin{ruledtabular}
    \begin{tabular}{c c c c c c c c c c c}
	    & $a$ (\AA) & $b$ (\AA) &  \textit{l$_1$}	&  \textit{l$_2$}	&  \textit{l$_3$}	& \textit{l$_4$}	& \textit{l$_5$} & \textit{l$_6$} & \textit{l$_7$} & \textit{l$_8$} \\ 
	\hline
     WTe$_2$ & 3.498 	& 6.338  &	 2.85 &	2.72 &	2.74 &	2.84 &	2.84 &	2.74 &	2.73 &	2.85 \\
     WSeTe   & 3.406	& 6.154  &   2.80 &	2.72 &	2.73 &	2.83 &	2.64 &	2.58 &	2.54 &	2.72 \\ 
     WSTe    & 3.344    & 6.064  &   2.79 &	2.72 &	2.73 &	2.83 &	2.50 &	2.46 &	2.41 &	2.63 \\
    \end{tabular}
    \end{ruledtabular}
    \caption{\textbf{Structural parameters of WTe$_2$ and associated Janus structures}. The lattice parameters $a$ and $b$ reduce on substitution of Te with Se/S. The different W-X bond lengths are labelled \textit{l$_1$--l$_8$} [see Figure~\ref{fig:structure}(b)]. All bond lengths are given in \AA.}
    \label{tab:structural_param}
\end{table*}


\subsection{\label{sec:electronic} Electronic and topological characterisation}

In this section, we begin by describing the electronic properties of the Janus monolayers and the associated topological characterisation.
Figure~\ref{fig:structure} shows the structure of pristine 1T' WTe$_2$, the Janus monolayers and their respective electronic band structures.
The top layer Te atoms of monolayer WTe$_2$ were replaced with its lighter congeners -- Se and S -- thereby constructing 1T' Janus monolayers WSeTe and WSTe.
We also carried out calculations of the formation energy to compare the stability of the two Janus structures. The formation energy, $E_{form}$, is given by $E_{form} = (E_{total} - E_W\times N_W - E_X\times N_X - E_Y\times N_Y)/N$. Here $E_{total}$ is the total energy of the Janus WXY monolayer. $E_W$, $E_X$, $E_Y$ are the energies of W, X and Y atoms in their respective bulk phases, $N_W$, $N_X$, $N_Y$ are the number of each of the atoms present in the unit cell and $N$ is the number of formula units present. We find that the formation energy of WSeTe is $\sim -0.52$ eV/f.u., while that of WSTe is $\sim -0.60$ eV/f.u. Therefore, the formation of WSTe is energetically more favourable than WSeTe.
The structural parameters associated with the pristine WTe$_2$ and the derived Janus structures are given in Table~\ref{tab:structural_param}.
The band structure of both WSeTe and WSTe reveal extensive splitting of band degeneracy due to the strongly broken out-of-plane symmetry, when compared to WTe$_2$.
We find that all three materials have non-zero direct bandgap at every point in the Brillouin zone.
The electron and the hole pockets present in WTe$_2$ along the high symmetry direction X-$\Gamma$ vanish for both the Janus monolayers leaving the systems gapped, but the band splitting leads to a reduced band gap.
Along the same direction, the band gap decreases from 143 meV in monolayer WTe$_2$ to 32 meV in WSeTe and 6 meV in WSTe. As we shall see, this reduction in gap has important consequences for Berry curvature in these systems. \\

\begin{figure*}
\begin{center}
  \includegraphics[scale=0.55]{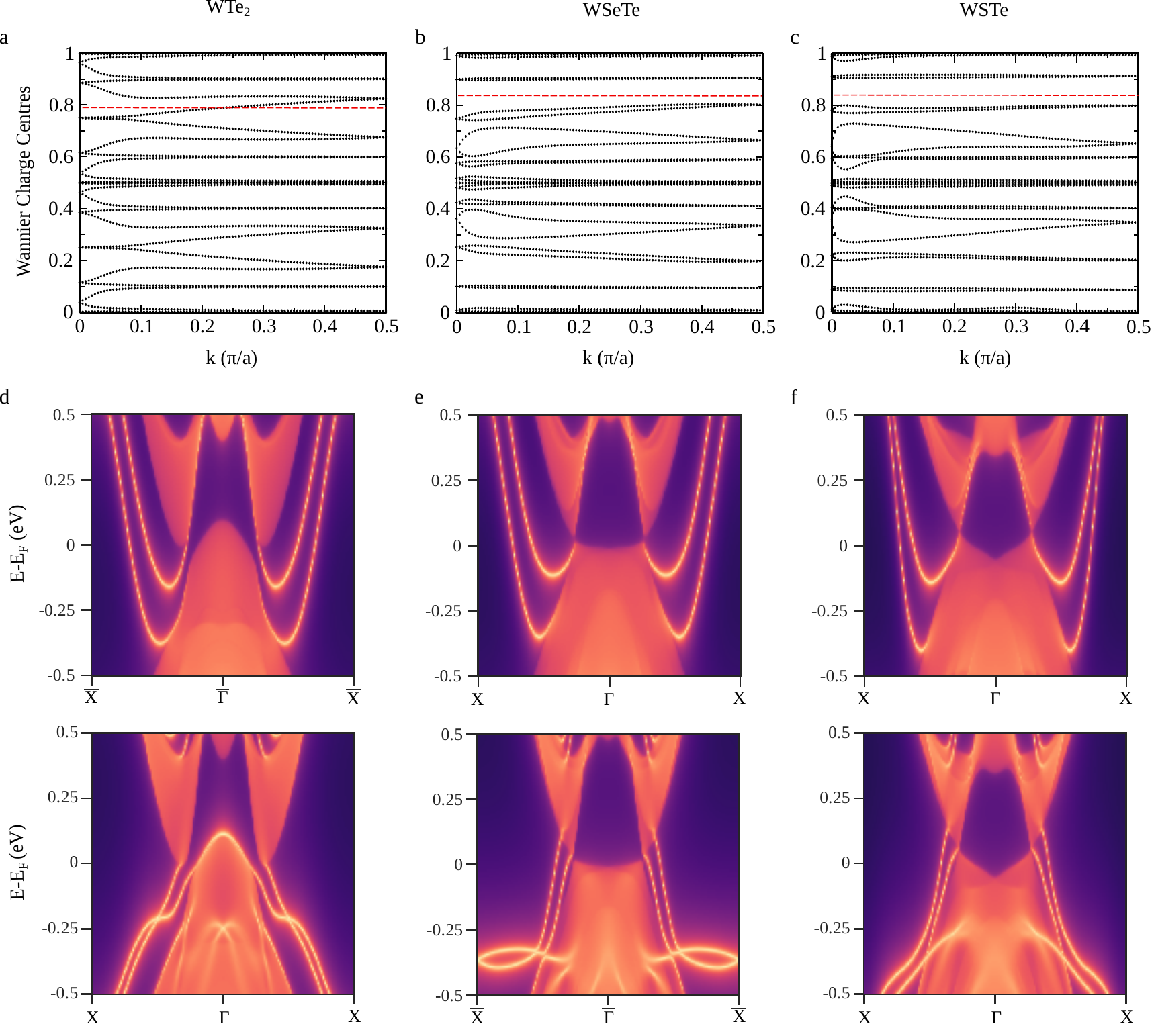}
   \caption{\textbf{Topological characterisation and edge states}. The Wannier Charge Centres (WCCs) tracked along the $k_z = 0$ plane for (a) WTe$_2$, (b) WSeTe and (c) WSTe. The line (red dashed) drawn parallel to the momentum axis intersects the WCCs (black dotted) at exactly 1, 0 and 0 times for WTe$_2$, WSeTe and WSTe, respectively, indicating that the pristine monolayer is topologically non-trivial while both the Janus structures are topologically trivial materials. (d-f) The edge density of states (DOS) associated with the three structures, along the two different types of edges present on construction of a one-dimensional ribbon. Brighter bands indicate higher DOS along the edge. The nature of the edge states is different in the Janus monolayers compared to pristine WTe$_2$, a manifestation of the different underlying topological properties.} \label{fig:topo_char}
\end{center}
\end{figure*}

Employing the methods described in Section~\ref{sec:methods}, we calculated the Z$_2$ topological invariant for the three monolayers. 
Figure~\ref{fig:topo_char} (a)-(c) present the WCCs along the TR invariant plane $k_z =0$.
The topological nature of the material can be deduced by drawing a line parallel to the $k$ axis.
For a two-dimensional material, if any such line intersects the WCCs odd number of times, it can be classified as topologically non-trivial. 
If the intersection is such that the line crosses the WCCs even number of times, including zero crossings, the material is trivial.
In Figure~\ref{fig:topo_char} (a), the line (red dashed) drawn parallel to $k$ axis intersects the WCCs (black dotted) an odd number of times indicating that the pristine monolayer WTe$_2$ is topologically non-trivial, as expected and consistent with previous reports ~\cite{Qian@tmdcQSH_science}. 
However, as shown in Figure~\ref{fig:topo_char} (b) and (c), we notably find that there exist zero or even number of such intersections, meaning both the Janus monolayers WSeTe and WSTe are topologically trivial materials. \\

Computation of the respective edge spectrum further gives insight to this difference in properties.
Figure~\ref{fig:topo_char} (d)-(f) shows the density of states projected onto the edges of the monolayers.
The figure details two sets of edge states [(d)-(f) top and bottom panels] for each monolayer, corresponding to the two different (W terminated and chalcogen terminated) edges present on the one dimensional ribbon.
An initial look suggests the edge states on all the three monolayers to be similar in nature, with a pair of bands on one edge moving up, towards the conduction band manifold and another pair on the opposite edge moving down, towards the valence band manifold.
On closer examination, we find that the origin of the bands in the Janus monolayers is different from the pristine WTe$_2$.
In WTe$_2$, for each pair of edge state, one of them originates from the valence band, while the other from the conduction band, leaving the one-dimensional ribbon gapless.
In case of the Janus monolayers, WSeTe and WSTe, both the pairs on either of the edges arise from either the valence band or the conduction band manifolds. 
In both these cases, any type of perturbation can easily gap the system, and the edge states will not be robust as in the case of WTe$_2$.
This difference in the nature of origin of the surface states between the Janus structures and WTe$_2$ lead to their respective trivial and non-trivial topological character, as obtained from our topological invariant calculations too. \\


\subsection{\label{sec:BCD} Berry curvature dipole}

\begin{figure*}
\begin{center}
  \includegraphics[scale=0.75]{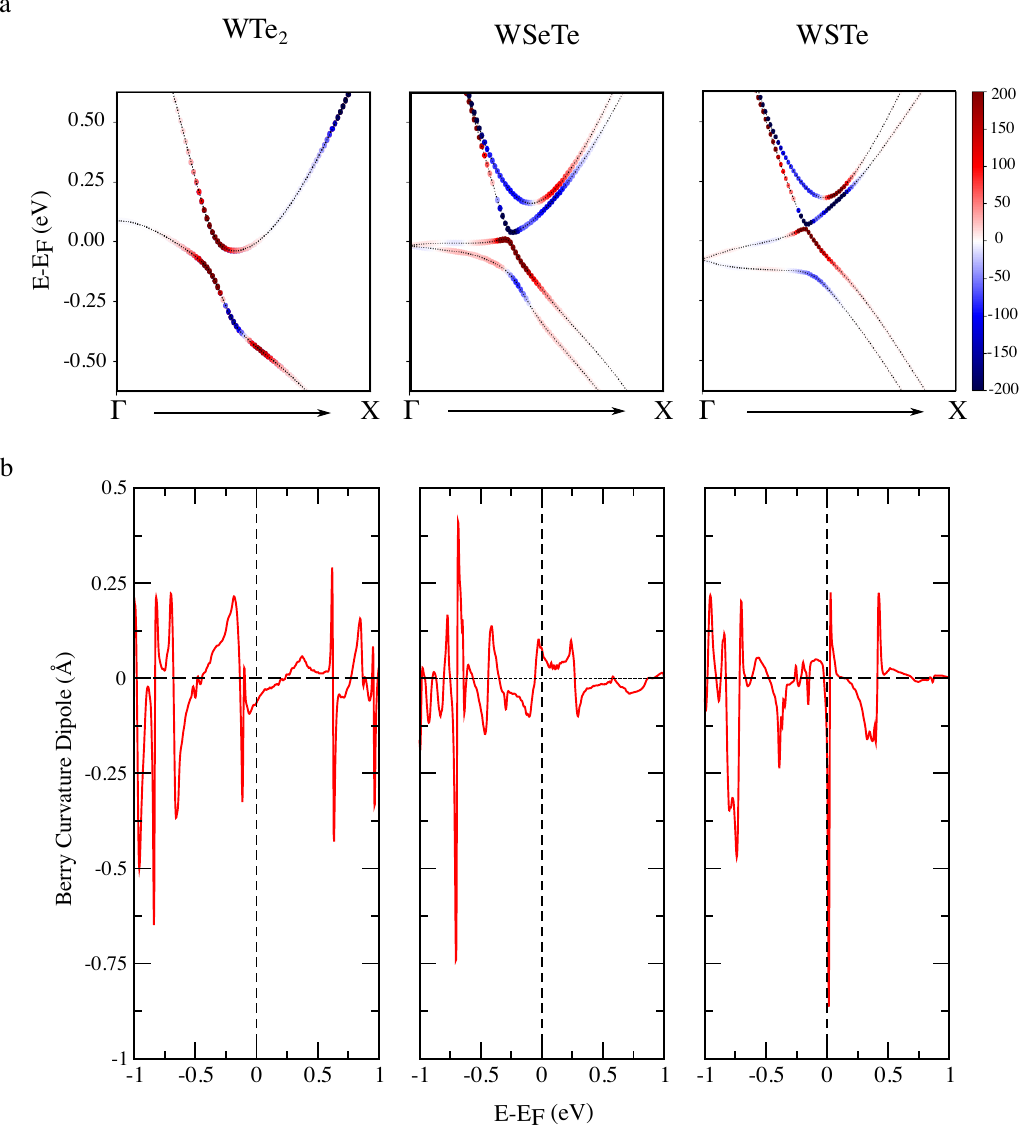}
   \caption{\textbf{Berry curvature and Berry curvature dipole from first-principles.} (a) Berry curvature along the high symmetry $\Gamma$-X direction, where the bandgap reduces as one layer of Te is replaced by its lighter congeners. The largest values of the Berry curvature occur near band anti-crossings. Here the red and blue colours indicate positive and negative values of the Berry curvature, respectively. (b) BCD associated with the monolayers as a function of the Fermi energy. Note the large peak in the vicinity of $E-E_{F}=0$ for WSTe. } \label{fig:bcd_dft}
\end{center}
\end{figure*}

As we discussed, recently, it has been found that systems with lowered symmetry and broken inversion display Hall-like currents in the non-linear regime, even when the TR symmetry is preserved~\cite{PhysRevLett.115.216806}.
Breaking inversion leads to a separation of positive and negative values of Berry curvature in the momentum space, resulting in a \textit{dipole}.
The 1T' structure of TMDC monolayers already breaks inversion, with its distorted octahedral metal-chalcogen coordination.
The only other symmetry present in these systems is a mirror plane along the $x$ axis ($M_x$), with the out-of-plane symmetry further strongly broken by the construction of Janus monolayers.
Such a system -- with broken inversion and lowered symmetry with a single mirror plane -- is ideal for the formation of BCD.
Since the Berry curvature only has an out-of-plane component and the BCD acts as a pseudo-vector in two-dimensions, as discussed in Section~\ref{sec:methods}, due to the presence of $M_x$ mirror plane, the only non-zero BCD component present is $D_{xz}$.

We next study the Berry curvature and BCD in our Janus structures. The distribution of Berry curvature along the high symmetry $\Gamma$-X direction, where the band gap closes due to substitution of the top layer chalcogen, is shown in Figure~\ref{fig:bcd_dft} (a). 
The red and blue colors indicate the positive and negative values respectively.
Berry curvature increases in intensity as the band gap is reduced, moving from Te to Se to S, taking up large values around points of anti-crossing. Also notably, the Berry curvature becomes significantly localized in momentum, specially near the Fermi level upon this elemental substitution.
We also highlight the strong splitting of the band degeneracy in the Janus structures and find that the resulting split-off bands carry opposite Berry curvatures.

Figure~\ref{fig:bcd_dft} (b) shows the non-zero $D_{xz}$ component of BCD as a function of the Fermi energy, in pristine WTe$_2$ and its associated Janus monolayers.
The BCD is a sensitive function of the Fermi level with peaks of varying intensities. We note that the BCD value at the Fermi level increases as one goes from Te to Se to S. This is a direct consequence of the increase in Berry curvature and its momentum space localization in the Janus structures.
The most noticeable feature is the sharp peak in BCD at the Fermi level for WSTe. The value of BCD at the Fermi level is nearly ten times larger in WSTe compared to WTe$_2$. Another intriguing feature occurs around $E-E_F=0.3$ eV, which is distinct to the Janus monolayers. This peak-dip structure -- with the BCD sharply changing sign -- is due to the splitting of the band degeneracy and creation of nearly touching bands with opposite signs of Berry curvature away from the Fermi level. 
Our obtained values of BCD and enhanced BCD around regions of band anti-crossings are consistent with previous results on TMDCs, with the dipole tunability realizable as a result of strain~\cite{son2019strain} and application of an electric field~\cite{zhang2018electrically}.
In comparison, twisted bilayer graphene displays a BCD two orders of magnitude higher than TMDCs. This higher value, when compared to TMDCs, has also been attributed to narrower bands and concentration of Berry curvature around the gapped Dirac cone in this system~\cite{pantaleon2021tunable}.
It would further be interesting to explore the strain-tuning of BCD in our Janus structures, since it has been shown that Berry curvature in TMDCs can be efficiently controlled using strain~\cite{wang2019janus,jena2019valley}.
Our calculations, therefore, demonstrate that Janus monolayers can be promising platforms to engineer large values of BCD and resulting non-linear Hall effects.  \\


\subsection{\label{sec:low_energy} Low-energy tilted massive Dirac model}

\begin{figure*}
\begin{center}
  \includegraphics[scale=0.75]{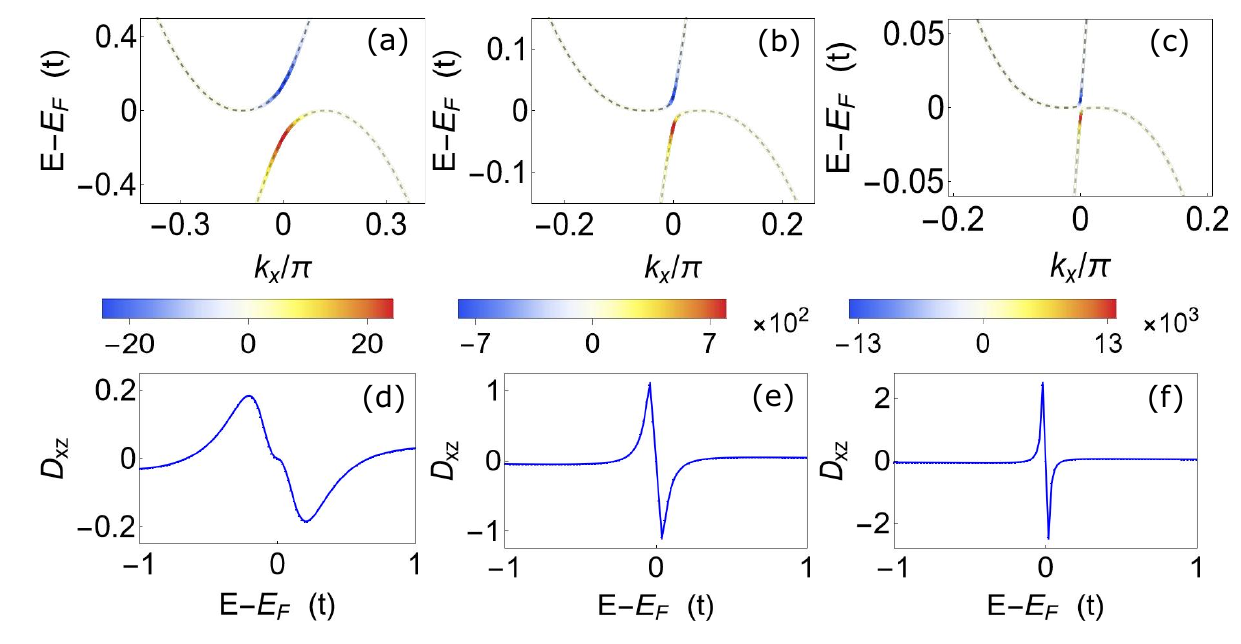}
   \caption{\textbf{Band Structure, Berry curvature and Berry curvature dipole in the tilted massive Dirac model.} Band structures for the tilted two-dimensional massive Dirac Hamiltonian along $k_x$ (for $k_y = 0$) for (a) $m =0.286$, (b) $m = 0.05$, (c) $m = 0.012$. Value of the Berry curvature is superimposed on band structure and corresponding color scales are shown. The corresponding BCD component $D_{xz}$ is plotted for (d) $m =0.286$, (e) $m = 0.05$, (f) $m = 0.012$. With decrease in $m$ the band gap decreases. The Berry curvature, which is concentrated at the band edges, increases and gets concentrated in smaller region with decrease in band gap. The BCD also increases with decreasing band gap due to higher value of the underlying Berry curvature. The BCD peaks get sharper and move closer to the Fermi energy with decreasing $m$. Other parameters are chosen to be $\eta = -1$, $v = 1$, $\alpha =1$, and $t=1$.} 
   \label{fig:band_bcd}
\end{center}
\end{figure*}

To gain further insights into our \textit{ab initio} results, we consider a simple low energy model for tilted two-dimensional massive Dirac cone. The Hamiltonian reads~\cite{PhysRevLett.121.266601}

\begin{equation}
    H = t k_x + v(k_y \sigma_x + \eta k_x \sigma_y) + (m/2 - \alpha k^2) \sigma_z,
    \label{eqn:low_energy_ham}
\end{equation}

where ($\sigma_x$, $\sigma_y$, $\sigma_z$) are the Pauli matrices, $v$ is the effective velocity, $t$ is the tilt of the Dirac cones along $k_x$ direction, ($k_x$, $k_y$) are the wave vectors, $k^2 = k_x^2+ k_y^2$, and $\eta = \pm 1$. Here $m$ is the mass parameter which controls the gap of the system. For regulating the topological properties as $k \rightarrow \infty$, the parameter $\alpha$ is introduced~\cite{shen_2012}. This Hamiltonian preserves TR symmetry, but inversion symmetry is broken by the addition of the tilt term with strength $t$. The dispersion relation is found to be

\begin{equation}
    E_{\pm} = t k_x \pm \sqrt{v^2 k^2 + (m/2 - \alpha k^2)^2}.
    \label{eqn:low_energy_dispersion}
\end{equation}

Note that the mass parameter, $m$, determines the gap and can be used to tune between our pristine and Janus monolayers. We can compute the Berry curvature, which has only a non-zero $z$ component, as

\begin{equation}
    \Omega^z_{\pm} = \pm \eta v^2 (m/2 - \alpha k^2)/ \{2[v^2 k^2 + (m/2 - \alpha k^2)^2]^{3/2}\}.
\end{equation}

As we discussed before, for our system the BCD tensor is a pseudo-vector in the $xy$ plane. In Figure~\ref{fig:band_bcd} (a)-(c), we show how the mass parameter, $m$, controls the band dispersion, at a fixed tilt. Here we choose three different values of $m$, which are in the ratio of the band gaps obtained from our first-principles calculations for WTe$_2$, WSeTe and WSTe. We find that with decreasing $m$, the band gap decreases as expected. As a consequence, the Berry curvature increases to very large values and gets concentrated in a smaller and smaller momentum region near the band edges. Note the different scales for the Berry curvature values in Figure~\ref{fig:band_bcd}(a) to (c).  

We next calculated the BCD using the above expression for the Berry curvature.  The variation of $D_{xz}$ with the Fermi level is shown in Figure~\ref{fig:band_bcd} (d)-(f) for the above mentioned $m$ values. We find that as $m$ is reduced, the maximum value of the BCD component increases sharply. The BCD peaks near the band edges where the Berry curvature is concentrated. However it is not exactly at the band edge since the group velocity ($v_i = \frac{\partial{E}}{\partial{k_i}}$) vanishes. Therefore, the peak occurs at a point that is close to the band edges where the product of Berry curvature and the group velocity is maximum. Furthermore, the peaks in the BCD become narrower and move closer to the $E-E_F=0$ with decreasing $m$, due to the Berry curvature becoming more localized in the momentum space. For the smallest values of the mass parameter ($m=0.012$), we find a large peak near $E-E_F=0$, closely resembling our DFT result for WSTe. It is remarkable that such a simple two-band effective model captures the essential features of our first-principles results.

\section{\label{sec:conclusion}Summary}

We investigated 1T' Janus monolayers derived from WTe$_2$ parent structure using a combination of density functional and tight-binding model computations. We discovered that the Janus monolayers of WSeTe and WSTe are in the topologically trivial phase, in contrast to the quantum spin Hall insulator WTe$_2$. We characterized their topology using a direct calculation of the Wannier charge centers as well as an examination of the edge states. Motivated by growing interest in non-linear Hall effects, we studied the Berry curvature dipole in these systems and found that all of them exhibit large values. A notable finding is a pronounced peak in the non-zero dipole component near the Fermi level for WSTe monolayer, resulting from a reduced band gap which in turn gives rise to a large Berry curvature. We further proposed a low-energy massive Dirac model, which captures the main features of our first-principles results. Our findings put forward Janus monolayers as promising platforms to engineer tunable non-linear Hall effects, and we hope that this motivates future studies of topology and Berry curvature dipole in Janus structures. \\

\section*{Acknowldegments}

We acknowledge discussions with D. Varghese, A. Bandyopadhyay, H. Liu, S. A. Yang,  S. Bhowal, N. A. Spaldin, N. P. Aetukuri and M. Jain. N.B.J. acknowledges support from the Prime Minister's Research Fellowship (PMRF). S. R. is supported by the Kishore Vaigyanik Protsahan Yojana (KVPY) fellowship. A.N. acknowledges support from the start-up grant (SG/MHRD-19-0001) of the Indian Institute of Science and DST-SERB (project number SRG/2020/000153).

\bibliography{References.bib}

\end{document}